\documentclass[a4paper]{jpconf}
\usepackage{graphicx}
\usepackage{amsmath,bm,dsfont,amssymb,amsfonts}

\renewcommand\[{\begin{equation}}
\renewcommand\]{\end{equation}}

\newcommand{\ssigma}{{\bm \sigma}}

\newcommand{\be}{\begin{equation}}
\newcommand{\ee}{\end{equation}}

\newcommand{\A}{{\cal A}}
\newcommand{\B}{{\cal B}}

\begin{document}
\title{A variational lower bound on the ground state of a many-body system and the squaring parametrization of density matrices}

\author{F. Uskov$^1$,  O. Lychkovskiy$^{1,2}$}

\address{$^1$ Skolkovo Institute of Science and Technology,	Nobel street 3, Moscow  121205, Russia}
\address{$^2$ Steklov Mathematical Institute of Russian Academy of Sciences,	Gubkina str., 8, Moscow 119991, Russia}

\ead{fel1992@mail.ru}

\begin{abstract}
A variational {\it upper} bound on the ground state energy $E_{\rm gs}$ of a quantum system, $E_{\rm gs} \leqslant \langle \Psi|H| \Psi \rangle$, is well-known (here $H$ is the Hamiltonian of the system and $\Psi$ is an arbitrary wave function). Much less known are variational {\it lower} bounds on the ground state. We consider one such bound which is valid for a many-body translation-invariant lattice system. Such a lattice can be divided into clusters which are identical up to translations. The Hamiltonian of such a system can be written as $H=\sum_{i=1}^M H_i$, where a term $H_i$ is supported on the $i$'th cluster. The bound reads $E_{\rm gs}\geqslant M \inf\limits_{\rho_{cl} \in {\mathbb S_{cl}^G}} \tr_{cl}\rho_{cl} \, H_{cl} $, where ${\mathbb S_{cl}^G}$ is some wisely chosen set of reduced density matrices of a single cluster. The implementation of this latter variational principle can be hampered by the difficulty of parameterizing the set  $\mathbb M$, which is a necessary prerequisite for a variational procedure. The root cause of this difficulty is the nonlinear positivity constraint $\rho>0$ which is to be satisfied by a density matrix. The squaring parametrization of the density matrix, $\rho=\tau^2/\tr\tau^2$, where $\tau$ is an arbitrary (not necessarily positive) Hermitian operator, accounts for positivity automatically. We discuss how the squaring parametrization can be utilized to find variational lower bounds on ground states of translation-invariant many-body systems. As an example, we consider a one-dimensional Heisenberg antiferromagnet.
\end{abstract}

\section{Introduction}
The ground state of a many-particle system is one of the central objects studied in condensed matter physics.
The ground state energy as a rule cannot be calculated exactly.
In strongly correlated systems, it is also difficult to apply the perturbation theory. A common way to asses the ground state energy is via variational methods. An upper bound on the ground state energy, $E_{\rm gs} \leqslant \langle \Psi|H| \Psi \rangle$, is well-known.
It is often desirable to supplement the latter with a {\it lower} bound.
Methods for obtaining lower bounds on ground state energies of many-body systems exist \cite{Anderson,NishimoriOzeki,MattisPan,Mazziotti,Baumgratz}, but they are much less developed than standard variational methods.
In this paper we suggest  one such method applicable to translation-invariant lattice systems with local interactions. The method is applied to a simple system, and its merits and prospects are discussed.

\section{Lower bound on the ground state energy of a translation-invariant lattice system}

\subsection{Our lower bound}

\begin{figure}[t]
	\begin{minipage}{7pc}
		\includegraphics[width=7pc]{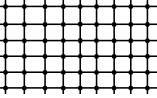}
	\end{minipage}\hspace{2pc}%
	\begin{minipage}{7pc}
		\caption{\label{label} Square lattice.}
	\end{minipage}\hspace{2pc}%
	\begin{minipage}{7pc}
		\includegraphics[width=7pc]{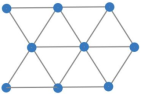}
	\end{minipage}\hspace{2pc}%
	\begin{minipage}{7pc}
		\caption{\label{label} Triangular  lattice.}
	\end{minipage}
\end{figure}
We consider a system of spins on a lattice with $N$ sites. The lattice is invariant with respect to the group of translation and, for two- and three-dimensional lattices, rotations. For example, this can be a linear chain in one dimension, a square or a triangular lattice in two dimensions (see figures 1, 2) {\it etc}.  Due to the symmetry, a lattice can be divided into identical clusters. The Hamiltonian of the system is defined on the lattice and is invariant under a group $G$ which contains the symmetries of the lattice and, in general, some other symmetries,
\be
U\,H\,U^\dagger =H~~~~~\forall ~~U\in G.
\ee
The Hamiltonian  can be written as
\be\label{H}
H=\sum_{i=1}^M H_i,
\ee
where $H_i$ is the local Hamiltonian of the $i$'th cluster, and the total number of clusters is $M$. In what follows we will use a special notation $H_{cl}\equiv H_1$ for the first cluster. Local terms $H_i$ can be transformed one to another by group actions, i.e.
\be\label{Hi}
H_i=U\,H_j\,U^\dagger~~~~{\rm for~ some}~~~U\in G.
\ee

Now let us derive our variational lower bound. It is well known that $E_{gs} =
\inf\limits_{ \Psi}\langle \Psi | H |\Psi\rangle$, where infimum is taken over all normalized vectors $\Psi$ of the Hilbert space.  We observe that this equality can be alternatively formulated in terms of density matrices  $\rho$ of the closed quantum system under considerations. Remind that a density matrix is an operator which satisfies three conditions,
\[{\rho ^ \dagger } = \rho ;\qquad \mathop{\rm{tr}}\rho  = 1;\qquad  \rho  > 0. \label{conditions on rho}\]
One can replace optimization over vectors $\Psi$ by optimization over density matrices $\rho$:
\[\label{inf over rho}
E_{gs} =
\inf\limits_{ \rho \in {\mathbb S}^G} \mathop{\rm{tr}} H \rho,
\]
Here  ${\mathbb S}^G$ is a set of density matrices $\rho$  invariant under the group $G$, i.e. in addition to conditions~\eqref{conditions on rho} the density matrices from the set ${\mathbb S}^G$ satisfy
\be\label{G-invariant rho}
U\,\rho\,U^\dagger=\rho~~~~{\rm for~ all}~~~U\in G.
\ee
Indeed, the density matrix which saturates the infimum in eq. \eqref{inf over rho} is simply the normalized projection onto the ground state subspace. This observation justifies eq. \eqref{inf over rho}. It should be stressed that density matrices appear here and in what follows as a formal tool for calculating a bound on the ground state of a {\it closed} system. Whether a system can be actually prepared in mixed states described by these density matrices is of no relevance in our argument.

 Further, using eqs. \eqref{H} and \eqref{Hi}, we get
\[\label{Egs 2}
E_{gs} =
M \inf\limits_{ \rho \in {\mathbb S}^G} \tr_{cl} \left( H_{cl} \tr_{\overline{cl}}\rho\right),
\]
where $\tr_{cl}$ and $\tr_{\overline{cl}}$ are partial traces over the cluster and its complement, respectively. Note that while $\rho_{cl}\equiv\tr_{\overline{cl}}\rho$ is just the reduced density matrix of the cluster, variation in eq. \eqref{Egs 2} is not performed over the set of {\it all} $\rho_{cl}$. Instead, the minimization is  performed over those $\rho_{cl}$ which can be obtained from $\rho \in {\mathbb S}^G$. The set of $\rho_{cl}$ satisfying the latter condition is unknown. However, we can lower bound $E_{gs}$ by performing minimization over a larger set ${\mathbb S}_{cl}^G$ of the reduced density matrices of the cluster symmetric under the group $G$. This way we obtain our variational lower bound
\[\label{our bound}
E_{gs} \geq
M \inf\limits_{ \rho_{cl} \in {\mathbb S}_{cl}^G} \tr_{cl}  H_{cl} \rho_{cl}.
\]
This bound is the main general result of the present paper. It should be stressed that $H_{cl}$ is {\it not} invariant under  the group $G$, in contrast to $\rho_{cl} \in {\mathbb S}_{cl}^G$.

We further observe that this bound can be enhanced by requiring that $\rho_{cl}$ satisfies local sum rules which follow from the anti-Hermitian Stationary Schr\"odinger equation \cite{Shpagina}.

\subsection{Density matrix parametrization}
In order to be able to perform minimization in eq. \eqref{our bound} one needs to parameterize the set ${\mathbb S}_{cl}^G$ of density matrices. 
Among the conditions which determine this set, the positivity condition, $\rho_{cl}>0$, is the most problematic due to its nonlinear nature. A squaring parametrization has been developed in \cite{SqParam} which automatically accounts for the positivity as well as other  conditions of the type \eqref{conditions on rho},\eqref{G-invariant rho}. In addition, it is well-suited for many-body systems. Its main idea is that if we take an arbitrary hermitian matrix $\tau$  and square and normalize it, we get a valid density matrix:
\[\rho  = \frac{{{\tau ^2}}}{{\mathop{\rm{tr}}{\tau ^2}}},~~~~~~{\rm where}~~~~~ {\tau ^ \dagger } = \tau. \label{tau}\]
We will use the squaring parametrization of density matrices to practically apply the bound \eqref{our bound}.

\subsection{Comparison to the Anderson bound}

Let us remind the Anderson bound, which is arguably the first and the most widely used lower bound on $E_{gs}$ \cite{Anderson}. It is based on a simple fact that an infimum of a sum is greater than the sum of infima. This leads to the bound
\be
E_{gs} \geq
M \inf\limits_{ \rho_{cl} } \, \tr_{cl}  H_{cl} \rho_{cl}.
\ee
The infimum here is taken over all density matrices $\rho_{cl}$ of a cluster. For this reason, the Anderson bound is weaker than our bound \eqref{our bound}.

\section{Application to a system of spins $1/2$ with Heisenberg interactions}

\subsection{Spin systems with Heisenberg interactions}

In the present section we demonstrate how the bound \eqref{our bound} can be applied to  a system of $N$ spins with the Heisenberg interaction. The Hamiltonian of this system reads
\begin{equation}\label{H Heisenberg}
	H = \sum_{<i,j>} \left( {{{\bm \ssigma}_i}{{\bm \ssigma}_j}} \right) ,
\end{equation}
where  the sum is taken over all neighbouring sites on the lattice, ${\bm \ssigma}_i$ is the vector consisting of three Pauli matrices of the $i$'th spin and $\left( {{{\bm \ssigma}_i}{{\bm \ssigma}_j}} \right)$ is the corresponding scalar product of sigma-matrices. This Hamiltonian is invariant with respect to a global $SU(2)$ symmetry, in other words, to the simultaneous rotations of all spins. It is also invariant under inversion of time and respects the spatial symmetries of the lattice.

As is discussed in details in \cite{SqParam,Shpagina}, a density matrix of the system \eqref{H Heisenberg} invariant under $SU(2)$ rotations and time inversion  can be expressed in terms of scalar products of sigma matrices; the same is true for  the auxiliary matrix  $\tau$ from eq. \eqref{tau}:
\def\arraystretch{1.8}
\begin{align}
\rho  = 2^{-N} \sum_\A a_\A A_\A,\qquad \qquad  &\tau  = \sum_\A b_\A A_\A,
\\
\{ A_\A \}  = \{ 1,  \;\;({\bf \ssigma}_j{\bf\ssigma}_k), \;\;
({\bf \ssigma}_j{\bf\ssigma}_k)({\bf \ssigma}_l{\bf\ssigma}_m) & \;\;,\;\;...\},~~i,k,l,m,...=1,2,...N.\label{set}
\end{align}
Here $\A$ is a multi-index which enumerates the set \eqref{set}, $b_i$ are arbitrary real numbers while $a_i$ are some functions of $b_i$ determined by eq. \eqref{tau}.

\subsection{Properties  of the set $\{A_\A\}$}
Here we discuss some properties  of the set $A_\A$ and a related set  $B_\A$ (see below). Some of this properties have immediate consequences for the implementation of our variational lower bound, while others may prove useful in further developments.

 First, we list useful relations \cite{SqParam}
\[\begin{array}{l}
{({{\ssigma}_1}{{\ssigma}_2})^2}\qquad \quad {\mkern 1mu} {\mkern 1mu}  = \;\;{\mkern 1mu} 3 - 2({{\ssigma}_1}{{\ssigma}_2})\\
({{\ssigma}_1}{{\ssigma}_2})({{\ssigma}_2}{{\ssigma}_3})\;\;\;\; = \;\;\;({{\ssigma}_1}{{\ssigma}_3}) - i({{\ssigma}_1}{{\ssigma}_2}{{\ssigma}_3})\\
({{\ssigma}_1}{{\ssigma}_2})({{\ssigma}_1}{{\ssigma}_2}{{\ssigma}_3}) =  - ({{\ssigma}_1}{{\ssigma}_2}{{\ssigma}_3}) - 2i({{\ssigma}_1}{{\ssigma}_3}) + 2i({{\ssigma}_2}{{\ssigma}_3})\\
({{\ssigma}_1}{{\ssigma}_2}{{\ssigma}_3})({{\ssigma}_1}{{\ssigma}_2}) =  - ({{\ssigma}_1}{{\ssigma}_2}{{\ssigma}_3}) + 2i({{\ssigma}_1}{{\ssigma}_3}) - 2i({{\ssigma}_2}{{\ssigma}_3})\\
({{\ssigma}_1}{{\ssigma}_2})({{\ssigma}_2}{{\ssigma}_3}{{\ssigma}_4}) = ({{\ssigma}_1}{{\ssigma}_3}{{\ssigma}_4}) - i({{\ssigma}_1}{{\ssigma}_3})({{\ssigma}_2}{{\ssigma}_4}) + i({{\ssigma}_1}{{\ssigma}_4})({{\ssigma}_2}{{\ssigma}_3})\\
({{\ssigma}_2}{{\ssigma}_3}{{\ssigma}_4})({{\ssigma}_1}{{\ssigma}_2}) = ({{\ssigma}_1}{{\ssigma}_3}{{\ssigma}_4}) + i({{\ssigma}_1}{{\ssigma}_3})({{\ssigma}_2}{{\ssigma}_4}) - i({{\ssigma}_1}{{\ssigma}_4})({{\ssigma}_2}{{\ssigma}_3})\\
{({{\ssigma}_1}{{\ssigma}_2}{{\ssigma}_3})^2}{\mkern 1mu} {\mkern 1mu} {\mkern 1mu} \qquad {\mkern 1mu}  = 6 - 2({{\ssigma}_1}{{\ssigma}_2}) - 2({{\ssigma}_1}{{\ssigma}_3}) - 2({{\ssigma}_2}{{\ssigma}_3})\\
({{\ssigma}_1}{{\ssigma}_2}{{\ssigma}_3})({{\ssigma}_1}{{\ssigma}_2}{{\ssigma}_4}) =  - ({{\ssigma}_1}{{\ssigma}_3})({{\ssigma}_2}{{\ssigma}_4}) - ({{\ssigma}_1}{{\ssigma}_4})({{\ssigma}_2}{{\ssigma}_3}) + 2({{\ssigma}_3}{{\ssigma}_4}) + i({{\ssigma}_1}{{\ssigma}_3}{{\ssigma}_4}) + i({{\ssigma}_2}{{\ssigma}_3}{{\ssigma}_4})\\
({{\ssigma}_1}{{\ssigma}_2}{{\ssigma}_3})({{\ssigma}_1}{{\ssigma}_4}{{\ssigma}_5}) =  + ({{\ssigma}_2}{{\ssigma}_4})({{\ssigma}_3}{{\ssigma}_5}) - ({{\ssigma}_2}{{\ssigma}_5})({{\ssigma}_3}{{\ssigma}_4}) - i({{\ssigma}_1}{{\ssigma}_2})({{\ssigma}_3}{{\ssigma}_4}{{\ssigma}_5}) + i({{\ssigma}_1}{{\ssigma}_3})({{\ssigma}_2}{{\ssigma}_4}{{\ssigma}_5}),
\end{array}\]
where $({{\ssigma}_1}{{\ssigma}_2}{{\ssigma}_3})$ is the mixed product of sigma matrices. Further, a product of two mixed products can always be represented through scalar products:
\[{\rm{(}}{{\ssigma}_1}{{\ssigma}_2}{{\ssigma}_3}{\rm{)(}}{{\ssigma}_4}{{\ssigma}_5}{{\ssigma}_6}{\rm{) = }}\det \left( {\begin{array}{*{20}{c}}
	{{\rm{(}}{{\ssigma}_1}{{\ssigma}_4}{\rm{)}}}&{{\rm{(}}{{\ssigma}_2}{{\ssigma}_4}{\rm{)}}}&{{\rm{(}}{{\ssigma}_3}{{\ssigma}_4}{\rm{)}}}\\
	{{\rm{(}}{{\ssigma}_1}{{\ssigma}_5}{\rm{)}}}&{{\rm{(}}{{\ssigma}_2}{{\ssigma}_5}{\rm{)}}}&{{\rm{(}}{{\ssigma}_3}{{\ssigma}_5}{\rm{)}}}\\
	{{\rm{(}}{{\ssigma}_1}{{\ssigma}_6}{\rm{)}}}&{{\rm{(}}{{\ssigma}_2}{{\ssigma}_6}{\rm{)}}}&{{\rm{(}}{{\ssigma}_3}{{\ssigma}_6}{\rm{)}}}
	\end{array}} \right)
\label{mixeddet}
.\]

One can introduce scalar product on the space of operators according to
\[(X,Y) \equiv \mathop{\rm{tr}}(X^\dagger Y) \]
(not to be confused with the scalar product of sigma matrices).
We consider the case when $X,Y\in \{A_\A\}.$ If supports of $X$ and $Y$ on a lattice do not coincide, then $(X,Y) = 0$.
If $X$ and $Y$ have the same support, then $\left( {X,Y} \right) = {2^N}{3^C}$,
where  $C$ is the number of cycles  arising when the bonds contained in $X$ are superimposed on the bonds contained in $Y$ ({\it cf.} ref. \cite{BeachSandvik}). In particular,
\[{\rm{tr}}(\;({{\ssigma}_i}{{\ssigma}_j})({{\ssigma}_j}{{\ssigma}_k})...({{\ssigma}_l}{{\ssigma}_m})({{\ssigma}_m}{{\ssigma}_i})\;) = 3 \cdot {2^N}\]
(one cycle) and
\[
{\rm{tr}}(\;({{\ssigma}_i}{{\ssigma}_j}{{\ssigma}_k})({{\ssigma}_i}{{\ssigma}_j}{{\ssigma}_n})\;({{\ssigma}_k}{{\ssigma}_l})...({{\ssigma}_m}{{\ssigma}_n})\;) =
{\rm{tr}}(\;({{\ssigma}_i}{{\ssigma}_j}{{\ssigma}_k})({{\ssigma}_i}{{\ssigma}_j}{{\ssigma}_k})\;) = 6 \cdot {2^N}.
\]

\begin{table}[t]
	\caption{\label{KN}The size $K(N)$ of the overcomplete set $\{A_\A\}$. One can see that for $N\lesssim 30$ this size is below the number of real parameters of the corresponding density matrix. }
	\begin{center}
		\begin{tabular}{lllllllllllllllllllllllllllllll}
			\br
			$N$&2 &3 &4 &5 &10 &15 &20 &30 &40 &50 &60 \\
			\mr
			$K(N)$&1 &3 &9 &25 &9495 &1E+7 &2E+10 &6E+17 &7E+25 &2E+34 &2E+43 \\
			$4^N$&16&64&256&1024&1048576&1E+9&1E+12&1E+18&1E+24&1E+30&1E+36\\
			\br
		\end{tabular}
	\end{center}
\end{table}

Let us consider a set  ${B_\B}$ of the following form:
$$
\{ B_\B \}  = \{ ({\bf \ssigma}_p{\bf\ssigma}_r{\bf\ssigma}_s),\;\;
({\bf \ssigma}_p{\bf\ssigma}_r{\bf\ssigma}_s)({\bf \ssigma}_j{\bf\ssigma}_k), \;\;
({\bf \ssigma}_p{\bf\ssigma}_r{\bf\ssigma}_s)({\bf \ssigma}_j{\bf\ssigma}_k)({\bf \ssigma}_l{\bf\ssigma}_m)\;\;,\;\;...\}
$$

Elements of sets $\{A_\A\}$ and $\{A_\A,B_\B\}\equiv{A_\A}\bigcup{B_\A}$ are not mutually orthogonal. For example,
\[\begin{array}{l}
\qquad \qquad \quad {\rm{X = (}}{{\ssigma}_1}{{\ssigma}_2}{\rm{)(}}{{\ssigma}_3}{{\ssigma}_4}{\rm{)}}\\
\qquad \qquad \quad {\rm{Y = (}}{{\ssigma}_1}{{\ssigma}_3}{\rm{)(}}{{\ssigma}_2}{{\ssigma}_4}{\rm{)}}\\
\qquad \qquad \quad {\rm{Z = (}}{{\ssigma}_1}{{\ssigma}_4}{\rm{)(}}{{\ssigma}_2}{{\ssigma}_3}{\rm{)}}\\
g = \left( {\begin{array}{*{20}{c}}
	{(XX)}&{(XY)}&{(XZ)}\\
	{(YX)}&{(YY)}&{(YZ)}\\
	{(ZX)}&{(ZY)}&{(ZZ)}
	\end{array}} \right) = \left( {\begin{array}{*{20}{c}}
	9&3&3\\
	3&9&3\\
	3&3&9
	\end{array}} \right)
\end{array}\]

Further, the set  $\{A_\A\}$ (and, consequently, $\{A_\A,B_\B\}$) is overcomplete. However, we put forward a hypothesis that  all linear dependencies within  $\{A_\A,B_\B\}$ can be described by
\[{\rm{ + (}}{{\ssigma}_1}{{\ssigma}_2}{\rm{)(}}{{\ssigma}_3}{{\ssigma}_4}{{\ssigma}_5}{\rm{) - (}}{{\ssigma}_1}{{\ssigma}_3}{\rm{)(}}{{\ssigma}_2}{{\ssigma}_4}{{\ssigma}_5}{\rm{)}} + {\rm{(}}{{\ssigma}_1}{{\ssigma}_4}{\rm{)(}}{{\ssigma}_2}{{\ssigma}_3}{{\ssigma}_5}{\rm{) - (}}{{\ssigma}_1}{{\ssigma}_5}{\rm{)(}}{{\ssigma}_2}{{\ssigma}_3}{{\ssigma}_4}{\rm{)}} = 0
\label{basisdep}
\]
for elements with an odd number of spins, and by
\[\det \left( {\begin{array}{*{20}{c}}
	{{\rm{(}}{{\ssigma}_1}{{\ssigma}_5}{\rm{)}}}&{{\rm{(}}{{\ssigma}_2}{{\ssigma}_5}{\rm{)}}}&{{\rm{(}}{{\ssigma}_3}{{\ssigma}_5}{\rm{)}}}&{{\rm{(}}{{\ssigma}_4}{{\ssigma}_5}{\rm{)}}}\\
	{{\rm{(}}{{\ssigma}_1}{{\ssigma}_6}{\rm{)}}}&{{\rm{(}}{{\ssigma}_2}{{\ssigma}_6}{\rm{)}}}&{{\rm{(}}{{\ssigma}_3}{{\ssigma}_6}{\rm{)}}}&{{\rm{(}}{{\ssigma}_4}{{\ssigma}_6}{\rm{)}}}\\
	{{\rm{(}}{{\ssigma}_1}{{\ssigma}_7}{\rm{)}}}&{{\rm{(}}{{\ssigma}_2}{{\ssigma}_7}{\rm{)}}}&{{\rm{(}}{{\ssigma}_3}{{\ssigma}_7}{\rm{)}}}&{{\rm{(}}{{\ssigma}_4}{{\ssigma}_7}{\rm{)}}}\\
	{{\rm{(}}{{\ssigma}_1}{{\ssigma}_8}{\rm{)}}}&{{\rm{(}}{{\ssigma}_2}{{\ssigma}_8}{\rm{)}}}&{{\rm{(}}{{\ssigma}_3}{{\ssigma}_8}{\rm{)}}}&{{\rm{(}}{{\ssigma}_4}{{\ssigma}_8}{\rm{)}}}
	\end{array}} \right) = 0\]
for elements with an even number of spins. Observe that the latter formula is a consequence of eqs. \eqref{mixeddet} and \eqref{basisdep}.
We tested this hypothesis for up to 10 spins.

The number of elements in the basis $\{A_\A\}$ without taking into account overcompleteness is equal to
\[K(N)=\sum_{k=0}^{[N/2]}C_N^{2k}\,(2k-1)!!\,\,,\]
where $[N/2]$ is the integer part of $N/2$.
$K(N)$ grows faster then exponentially with $N$ (see table~\ref{KN}), however it can be rather small for $N\sim 10$.

\subsection{An example}

\begin{table}
	\caption{\label{result} Anderson bound compared to the bound \eqref{our bound} for a translation-invariant linear chain with the Heisenberg Hamiltonian \eqref{H Heisenberg linear}. Given are the lower bounds on the ground state energy per spin,  $E_{gs}/N$. The exact Bethe ansatz result reads $E_{gs}/N=1-4\log 2\simeq -1.77259.$}
	\begin{center}
		\begin{tabular}{lllllllllllllllllllllllllllllll}
			\br
			cluster size&Anderson bound& bound \eqref{our bound}\\
			\mr
			3 &-2.0     & -2.0\\
			4 &-2.1547  & -2.0\\
			5 &-1.9279 & -1.8685\\
			6 &-1.9947 & -1.8685\\
			7 &-1.8908 & -1.8255\\
			\br
		\end{tabular}
	\end{center}
\end{table}

Here we consider a one-dimensional lattice of $N$ spins $1/2$ with the nearest-neighbour Heisenberg Hamiltonian
\begin{equation}\label{H Heisenberg linear}
	H = \sum_{j=1}^N \left( {{{\bm \ssigma}_j}{{\bm \ssigma}_{j+1}}} \right),
\end{equation}
where $\ssigma_{N+1}\equiv\ssigma_1$ which ensures translation invariance. This system is integrable by means of Bethe ansatz and the ground state energy per spin in the thermodynamic limit reads $E_{gs}/N=1-4\log 2$ \cite{Bethe}. Our goal is to lower bound $E_{gs}/N$ by means of our method and to compare this bound to the exact value and to the Anderson bound~\cite{Anderson}.

We start from considering a cluster with 4 spins. Its Hamiltonian reads
\[{H_{cl}} = ({{\ssigma}_1}{{\ssigma}_2}) + ({{\ssigma}_2}{{\ssigma}_3}) + ({{\ssigma}_3}{{\ssigma}_4})\label{Hcl}\]
The basis supported by the cluster reads
\[\label{basis 4 spins}
\begin{array}{l}
\{ {A_k}\}  = \{ 1,\;({{\ssigma}_1}{{\ssigma}_2}),\;({{\ssigma}_1}{{\ssigma}_3}),\;({{\ssigma}_1}{{\ssigma}_4}),\;({{\ssigma}_2}{{\ssigma}_3}),\;({{\ssigma}_2}{{\ssigma}_4}),\;({{\ssigma}_3}{{\ssigma}_4}),\;\\
\;\;\;\;\;\;\;\;\;\;\;\;\;\;\;({{\ssigma}_1}{{\ssigma}_2})({{\ssigma}_3}{{\ssigma}_4}),\;
({{\ssigma}_1}{{\ssigma}_3})({{\ssigma}_2}{{\ssigma}_4}),\;({{\ssigma}_1}{{\ssigma}_4})({{\ssigma}_2}{{\ssigma}_3})\}, ~~~k=0,1,,...9.
\end{array}\]

We apply the squaring parametrization  \cite{SqParam} in the form
\be\label{squaring unnormalized}
\rho_{cl}  = {\tau ^2}.
\ee
Here the reduced density matrix, $\rho_{cl}$, and the axillary matrix, $\tau$, are expanded in the basis \eqref{basis 4 spins} as
\[
\rho_{cl}  = 2^{-4}{a_k}{A_k},~~~~~~~~~~~~~\tau  = {b_k}{A_k},
\]
where summation over repeating indices is implied. The normalization of $\rho_{cl}$ is imposed by the constraint
\[{a_0} = 1.\label{normalization 4 spins}\]
The squaring parametrization \eqref{squaring unnormalized} implies that each coefficient $a_k$ is a quadratic function of coefficients $b_k$.

Translational invariance implies
\[{a_{(1,2)}} = {a_{(2,3)}} = {a_{(3,4)}},\;\;{a_{(1,3)}} = {a_{(2,4)}}.\]
Here we use self-explanatory notations for indexes, e.g. $a_{(1,2)}$ is the coefficient in front of $(\ssigma_1 \ssigma_2)$.
The Hamiltonian \eqref{Hcl} of the cluster  is not translationally invariant, but it posses a remaining mirror symmetry. For this reason two of the above equalities can be satisfied seamlessly:
\[
{b_{(1,2)}} = {b_{(3,4)}},\;\;{b_{(1,3)}} = {b_{(2,4)}}\;~~~\Rightarrow~~~\;
{a_{(1,2)}} = {a_{(3,4)}},\;\;{a_{(1,3)}} = {a_{(2,4)}}.
\]
The condition
\[a_{(1,2)}=a_{(2,3)}\label{condition 4 spins}\]
remains and should be accounted for during optimization.

Finally we perform a numerical search for a minimum of $\tr H_{cl} \tau^2 $ with the constraints \eqref{normalization 4 spins} and \eqref{condition 4 spins}, which are equivalent to  $\tr  \tau^2 =1$ and $\tr  (\ssigma_1,\ssigma_2)\tau^2 =(\ssigma_2,\ssigma_3)\tau^2$, respectively. The resulting bound is presented in table \ref{result}, along with the analogous bounds with other cluster sizes. One can see that for a given cluster size the bound \eqref{our bound} outperforms the Anderson bound. The caveat here is that for a given cluster size the calculations for our bound \eqref{our bound} require much more resources than those for the Anderson bound.   Whether  the bound \eqref{our bound} is able to compete with the Anderson bound in practical numerical computations is a question open for future research.

\section{Summary}

We have derived a variational {\it lower} bound \eqref{our bound} on the ground state energy of a quantum system possessing symmetries. The variation should be performed over a certain set of reduced density matrices. Technically, the variation can be performed by means of the squaring parametrization of density matrices \cite{SqParam} which automatically satisfies the positivity condition.   We have discussed how this bound can be applied for translation-invariant spin systems with the Heisenberg interaction. Promising results has been obtained in a simple example of a linear chain, see table \ref{result}.


\section*{Acknowledgements.} We are grateful to E. Shpagina and N. Il'in for useful discussions. The work was supported by the Russian Science Foundation under the grant No. 17-71-20158�.


\end{document}